\documentclass[aps,prb,twocolumn,floats,floatfix,superscriptaddress,showpacs]{revtex4}
\usepackage{epsfig}
\usepackage{color}
\usepackage{bm}
\usepackage{latexsym}
\begin{document}
\newcommand{\hide}[1]{}
\newcommand{\tbox}[1]{\mbox{\tiny #1}}
\newcommand{\half}{\mbox{\small $\frac{1}{2}$}}
\newcommand{\sinc}{\mbox{sinc}}
\newcommand{\const}{\mbox{const}}
\newcommand{\trc}{\mbox{Tr}}
\newcommand{\intt}{\int\!\!\!\!\int }
\newcommand{\ointt}{\int\!\!\!\!\int\!\!\!\!\!\circ\ }
\newcommand{\eexp}{\mbox{e}^}
\newcommand{\bra}{\left\langle}
\newcommand{\ket}{\right\rangle}
\newcommand{\EPS} {\mbox{\LARGE $\epsilon$}}
\newcommand{\ar}{\mathsf r}
\newcommand{\im}{\mbox{Im}}
\newcommand{\re}{\mbox{Re}}
\newcommand{\bmsf}[1]{\bm{\mathsf{#1}}}
\definecolor{red}{rgb}{1,0.0,0.0}
\title{Scattering at the Anderson transition: Power--law banded random
matrix model}
\author{J. A. M\'endez-Berm\'udez}
\affiliation{Max-Planck-Institut f\"ur Dynamik und Selbstorganisation,
Bunsenstrasse 10, D-37073 G\"ottingen, Germany}
\affiliation{Department of Physics, Ben-Gurion University,
Beer-Sheva 84105, Israel}
\affiliation{Instituto de F\'{\i}sica, Universidad Aut\'onoma de Puebla,
Apartado Postal J-48, Puebla 72570, Mexico}
\author{I. Varga}
\affiliation{Elm\'eleti Fizika Tansz\'ek, Fizikai
Int\'ezet, Budapesti M\H uszaki \'es Gazdas\'agtudom\'anyi Egyetem,
H-1521 Budapest, Hungary}
\affiliation{Fachbereich Physik und Wissenschaftliches Zentrum f\"ur
Materialwissenschaften, Philipps Universit\"at Marburg, D-35032 Marburg,
Germany}
\date{\today}
\begin{abstract}
We analyze the scattering properties of a periodic one-dimensional
system at criticality represented by the so-called power-law banded
random matrix model at the metal insulator transition.
We focus on the scaling of Wigner delay times $\tau$ and resonance
widths $\Gamma$. We found that the typical values of $\tau$ and
$\Gamma$ (calculated as the geometric mean) scale with the system
size $L$ as $\tau^{\mbox{\tiny typ}}\propto L^{D_1}$ and
$\Gamma^{\mbox{\tiny typ}} \propto L^{-(2-D_2)}$, where $D_1$ is the
information dimension and $D_2$ is the
correlation dimension of eigenfunctions of the corresponding closed
system.
\end{abstract}
\maketitle
\section{Introduction}
The investigation of the disorder induced metal--insulator transition
(MIT) is an exciting and yet unsolved problem in condensed matter physics.
During the past almost fifty years of research many important features
have been cleared out using analytical, numerical, and experimental
techniques. Yet this problem continues to bring interesting and
unexpected novelties to daylight.~\cite{A58,AMPJ95,AKL91}

Many characteristics of disordered metals and those of insulators have
been understood over the past. The spectral fluctuations of a disordered
metal are well described by random matrix theory, while the fluctuations
of a system in the insulating regime follow the Poisson
statistics.~\cite{SSLS93,AS86}
The nature and the details of the MIT, on the other hand, still belong
to the most intensively studied problems. We know already that the spectral
fluctuations of a system right at the transition have properties merged
from the two extremes.~\cite{SSLS93} However, relations between the
spectral and eigenfunction fluctuations show that there is an intimate
connection between the two~\cite{CLS96} which is related to anomalous
wave packet spreading and diffusion.~\cite{HK99}

Recently, several works have been devoted to deepen our understanding
of the scattering properties of disordered systems by analyzing the
distribution of resonance widths and Wigner delay
times.~\cite{MK05,OF05,MFME06,OKG03,F03,TC99,KW02a,KW02b,WMK06}
Both distribution functions have been shown to be closely related to the
properties of the corresponding closed system, i.e., the fractality of the
eigenstates and the critical features of the MIT. In this respect, detailed
analysis have been performed for the three-dimensional (3D) Anderson
model~\cite{KW02a} and also for the power law band random matrix (PBRM)
model.~\cite{MFDQS96,ME00,KT00,EM00,V03,M00}
The PBRM model, which describes a one-dimensional (1D) sample with random
long-range hopping, has been found to provide many of the features of the
localization-delocalization transition present in the 3D Anderson model.

In the present work we look at the PBRM model more closely to see how
general are the findings presented in
Refs.~\onlinecite{MK05,KW02a}, and \onlinecite{WMK06}. Here, we find good
scaling of the distribution
functions of resonance widths and Wigner delay times. Moreover,
we state new scaling relations of their typical values (i.e., their
geometric mean) determined by the information dimension, $D_1$, and the
correlation dimension, $D_2$, of the eigenfunctions of the corresponding
closed system. We note that $D_2$ also governs the spreading of wave packets
in systems at the MIT.\cite{HK99}

Finally, we have to mention that in recent microwave experiments
the study of systems in which the resonance states resemble the
characteristics of those of our interest was reported.~\cite{SS90}
Therefore the direct verification of our results is expected to be
available in the near future.

The organization of this paper is as follows. In the next section we
describe the model we use and define the scattering setup. Section III
is devoted to the analysis of the coupling between sample and lead.
In Sec. IV we present our results for the distribution functions of
resonance widths and Wigner delay times.
Finally, Sec. V is left for conclusions.
\section{Model}
The isolated sample of length $L$ is represented by an $L\times L$ real
symmetric matrix whose entries are randomly drawn from a normal distribution
with zero mean, $\langle H_{ij}\rangle = 0$, and a variance depending on the
distance of the matrix element from the diagonal as
\begin{equation}
\label{pbrm}
\bra (H_{ij})^2\ket = {1\over 2}
{{\delta_{ij}+1}\over
1+\left[ \sin\left(\pi|i-j|/L\right)/(\pi b/L)\right]^{2\alpha}}\, \, ,
\end{equation}
where $b$ and $\alpha$ are parameters. In order to reduce finite size
effects this expression already incorporates periodic boundary
conditions, where Eq.~(\ref{pbrm}) is known as the periodic PBRM model.
Nevertheless, for sites far away from each other and the
border, i.e., $1\ll |i-j|\ll N$, the variance decays with a power law
\begin{equation}
\left\langle (H_{ij})^2\right\rangle
    \sim \left (\frac{b}{|i-j|}\right )^{2\alpha}.
\end{equation}
Field-theoretical considerations~\cite{MFDQS96,M00,KT00} and
detailed numerical investigations~\cite{EM00,V03} verified that the
PBRM model undergoes a transition at $\alpha=1$ from localized states for
$\alpha >1$ to delocalized states for $\alpha < 1$. This transition
shows all the key features of the Anderson MIT, including multifractality
of eigenfunctions and non-trivial spectral statistics at the critical point.
At the center of the spectral-band a theoretical estimation for the
multifractal dimensions, $D_q$, of the eigenfunction $\psi({\bf r})$
gives~\cite{ME00}
\begin{equation}
\label{Dq}
D_q =\left\{
\begin{array}{cc}
 4b {\widetilde \Gamma}(q-1/2)[\sqrt{\pi}(q-1){\widetilde \Gamma}(q-1)]^{-1},
 & b \ll 1 \nonumber\\
 1-q(2\pi b)^{-1}, & b\gg 1
\end{array}
\right.
\end{equation}
where ${\widetilde \Gamma}$ is the Gamma function. $D_q$ is defined
through the R\'enyi entropy~\cite{VP03} of the
participation number~\cite{V03,M00,FE95,W80}
\begin{equation}
\label{RE}
{\cal R}_q = {{1}\over {1-q}}\ln {\cal N}_q\propto D_q\ln L \ ,
\end{equation}
where the participation number is defined as
\begin{equation}
\label{PN}
{\cal N}_q = \left(\int \left|\psi({\bf r})\right|^{2q} d{\bf r}\right)^{-1}
\end{equation}
showing a nontrivial scaling with respect to the linear size of the
system $L$.
Thus model (\ref{pbrm}) possesses a line of critical points $b\in (0,\infty)$,
where the multifractal dimensions $D_q$ change with $b$. The value of
parameter $b$ describes the strength of disorder. In other words it
is related to the classical conductance through the
relation~\cite{KT00} $g_c=4\beta b$,
where $\beta$ stands for the global symmetry of the system:
$\beta=1$ describes the presence and $\beta=2$ the absence of time
reversal symmetry.

Among all dimensions, the information dimension $D_1$ and the correlation
dimension $D_2$ play a prominent role.\cite{HP83} However, $D_1$ is obtained
as the $q\to 1$ limit of Eq.~(\ref{RE}), being the scaling of the Shannon
entropy.~\cite{HP83}

\begin{figure}
\begin{center}
    \epsfxsize=8.4cm
    \leavevmode
    \epsffile{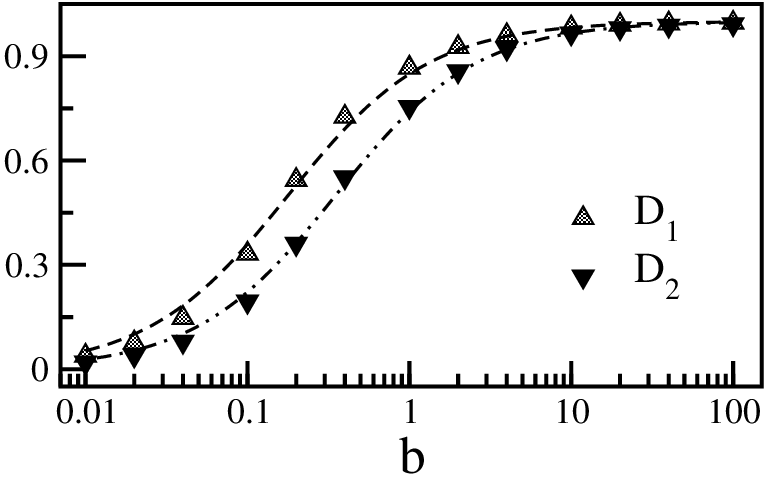}
\caption{$D_1$ and $D_2$ as a function of $b$. Lines
correspond to Eq.~(\ref{d2}) with $\sigma\approx 5.6$ (dashed) and
$\sigma\approx 2.85$ (dot-dashed). }
\label{fig:fig0}
\end{center}
\end{figure}

We have to mention that recently a phenomenological
formula was obtained in Ref.~\onlinecite{MKC05} for the exponent $D_2$
as a function of $b$ (and in this way as a function of $g_c$):
\begin{equation}
D(b)=\frac{1}{1+(\sigma b)^{-1}} \ ,
\label{d2}
\end{equation}
where $\sigma$ is a fitting parameter. For the present model we found
that $\sigma\approx 2.85$ reproduces well the numerical $D_2$ extracted
from ${\cal N}_2$,~\cite{V03} see Eq.~(\ref{PN}).
Similarly Eq.~(\ref{d2}) fits well the exponent
$D_1$ with $\sigma\approx 5.6$.  See Fig.~\ref{fig:fig0}.

We turn the isolated system to a scattering one by attaching one
semi-infinite single channel lead to it. The lead is described by the
following 1D tight-binding Hamiltonian:
\begin{equation}
\label{leads}
H_{\rm lead}=\sum_{n=1}^{-\infty} (|n\rangle\langle n+1|
                                 + |n+1\rangle\langle n|)\,\,.
\end{equation}

Using standard methods~\cite{MW69} one can write the scattering matrix
(a single complex number for one attached lead) in the form~\cite{KW02a,KW02b}
\begin{equation}
\label{smatrix}
S(E) = e^{i\Phi(E)}=1-2i \pi \, W^{\,T} \left (E{\bf 1}-{\cal H}_{\rm
eff}\right )^{-1}W,
\end{equation}
where ${\bf 1}$ is an $L\times L$ unit matrix
and ${\cal H}_{\rm eff}$ is an effective non-hermitian Hamiltonian given by
\begin{equation}
\label{Heff}
{\mathcal{H}}_{\rm eff}=H - i \pi WW^{\,T}.
\end{equation}
Here, $W$ is an $L\times 1$ vector with elements $W_n=w_0 \delta_{nn_0}$,
where $w_0$ is the coupling strength between sample and lead and $n_0$ is
the site at which the lead is attached. All our calculations take place
in an energy window close to the band center $(E=0)$. Then, the Wigner
delay time is given by~\cite{KW02a,KW02b,FS97}
\begin{equation}
\label{tauW}
\tau(E=0) = \left. {d\Phi(E)\over dE} \right|_{E=0}
   = \left. -2 \im \trc (E-{\cal H}_{\rm eff})^{-1}\right|_{E=0} \ .
\end{equation}
\begin{figure}
\begin{center}
    \epsfxsize=8.4cm
    \leavevmode
    \epsffile{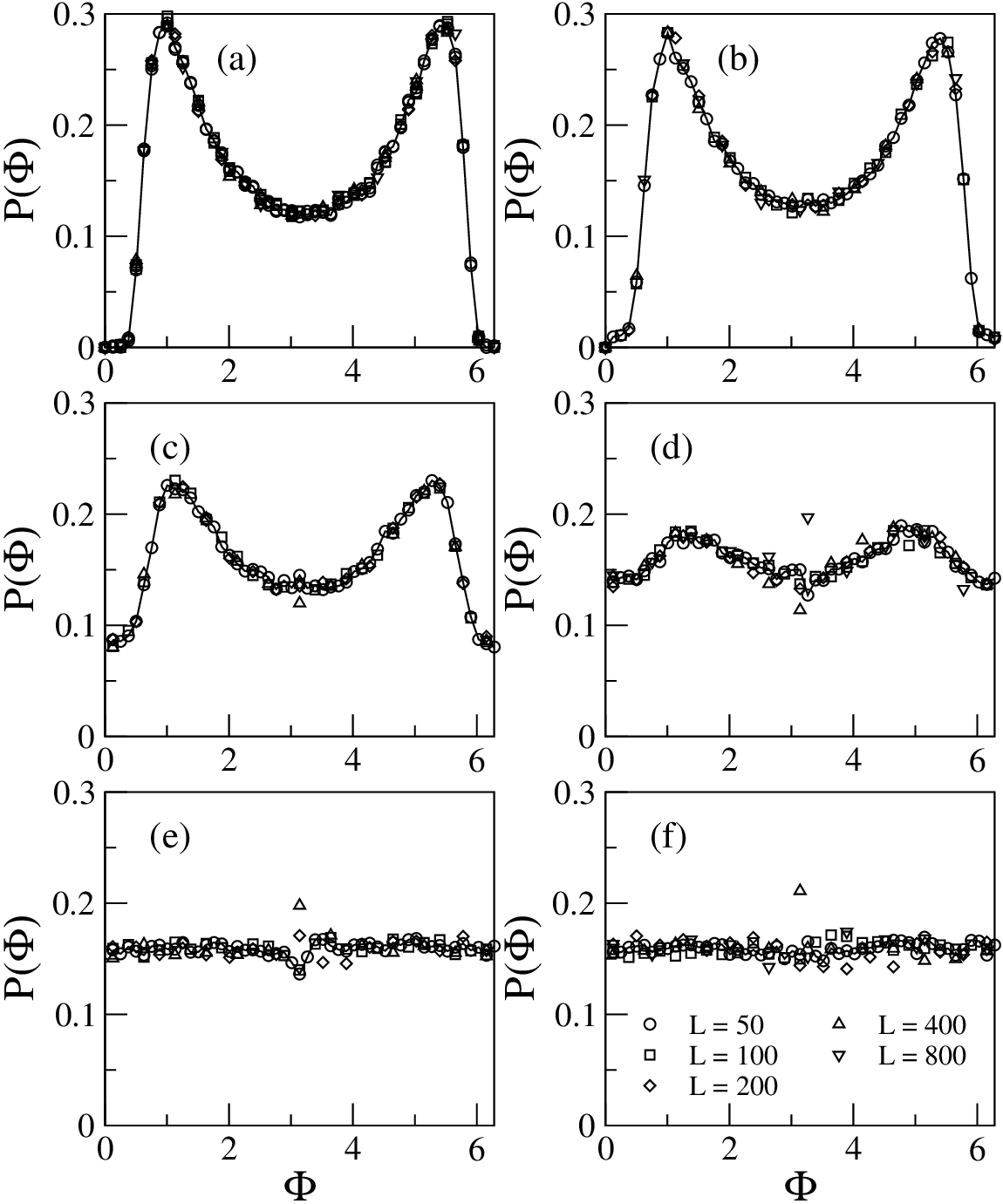}
\caption{${\cal P}(\Phi)$ corresponding to $w_0$ where $|\langle S\rangle|$
takes its minimum value (perfect coupling condition) for (a) $b=0.01$ and 0.04,
(b) $b=0.1$, (c) $b=0.4$, (d) $b=1$, (e) $b=4$, and (f) $b=10$.
We use (a) $w_0 = 0.6$, (b) $w_0 = 0.65$, (c) $w_0 = 0.9$, (d) $w_0 = 1.55$,
and (e) $w_0 = 3.4$. In (f) $w_0 = 5.1$, $w_0 = 5.43$, and $w_0 = 5.5$ for
$L=50$, $L=100$, and $L\ge 200$, respectively.}
\label{fig:fig1}
\end{center}
\end{figure}

In addition to the delay times, which captures the time-dependent aspects
of quantum scattering, the poles of the scattering matrix are also of
great relevance.~\cite{WMK06} They determine the conductance fluctuations
of a quantum dot in the Coulomb blockade regime~\cite{ABG02} or the current
relaxation.~\cite{AKL91} The poles of the scattering matrix show up as
resonances,
which are the complex eigenvalues ${\mathcal{E}}_n = E_n - i\Gamma_n/2$
of ${\mathcal{H}}_{\rm eff}$. $E_n$ and $\Gamma_n$ are the position and
width of the $n$th resonance, respectively. Of course for $w_0=0$ the real
part of the poles, $E_n$, correspond to the eigenvalues of the closed
system. The $\Gamma_n$s, on the other hand, are related to the lifetime
of the resonances through $\tau_n =1/\Gamma_n$. Thus, we may expect
close correspondence between Wigner delay times and the resonance
widths which, in fact, we calculate independently from each other.
Bellow we use matrices of sizes varying from $L=50$ up to $L=800$
to compute probability distribution functions of scattering phases
$\Phi$, Wigner delay times $\tau$, and resonance widths $\Gamma$.
For statistical processing a large number of disorder realizations
is used. In the case of ${\cal P}(\Phi)$ and ${\cal P}(\tau)$ we
have at least $100000$ data values. For the statistics of $\Gamma$
we use the eigenvalues around $E_n\sim 0$, about one tenth of the
total spectra. Here, at least $250000$ data values are used.

\section{Perfect coupling}

In this section we define more precisely the way the lead is
attached to the system under the condition of perfect coupling.
We should first note that the lead is attached to any site
(chosen at random) within the sample and that the ensemble average
also implies average over the position of the lead.
The value of the coupling strength, $w_0$, is defined by analyzing
the distribution of the scattering phases ${\cal P}(\Phi)$.
See also Ref.~\onlinecite{MK05}. In Fig.~\ref{fig:fig1} we present
${\cal P}(\Phi)$ for various values of $b$ and $L$. We use (here
and bellow) the coupling strengths $w_0$ such that
$|\langle S(w_0)\rangle|\approx 0$, where the corresponding
setup is associated with perfect coupling between sample and lead.
$\langle S\rangle$ is the ensemble average of the $S$ matrix.
Notice that for $b>1$ we recover the uniform distribution.
Moreover, in this case, the expression~\cite{OF05}
${\cal P}(\Phi) = 1/2\pi (\gamma + \sqrt{\gamma^2-1} \cos(\Phi))$
holds (not shown here); where
$\gamma = (1+|\langle S\rangle|^2)/(1-|\langle S\rangle|^2)$ and
$0 \le \langle S\rangle \le 1$.

\begin{figure}
\begin{center}
    \epsfxsize=8.4cm
    \leavevmode
    \epsffile{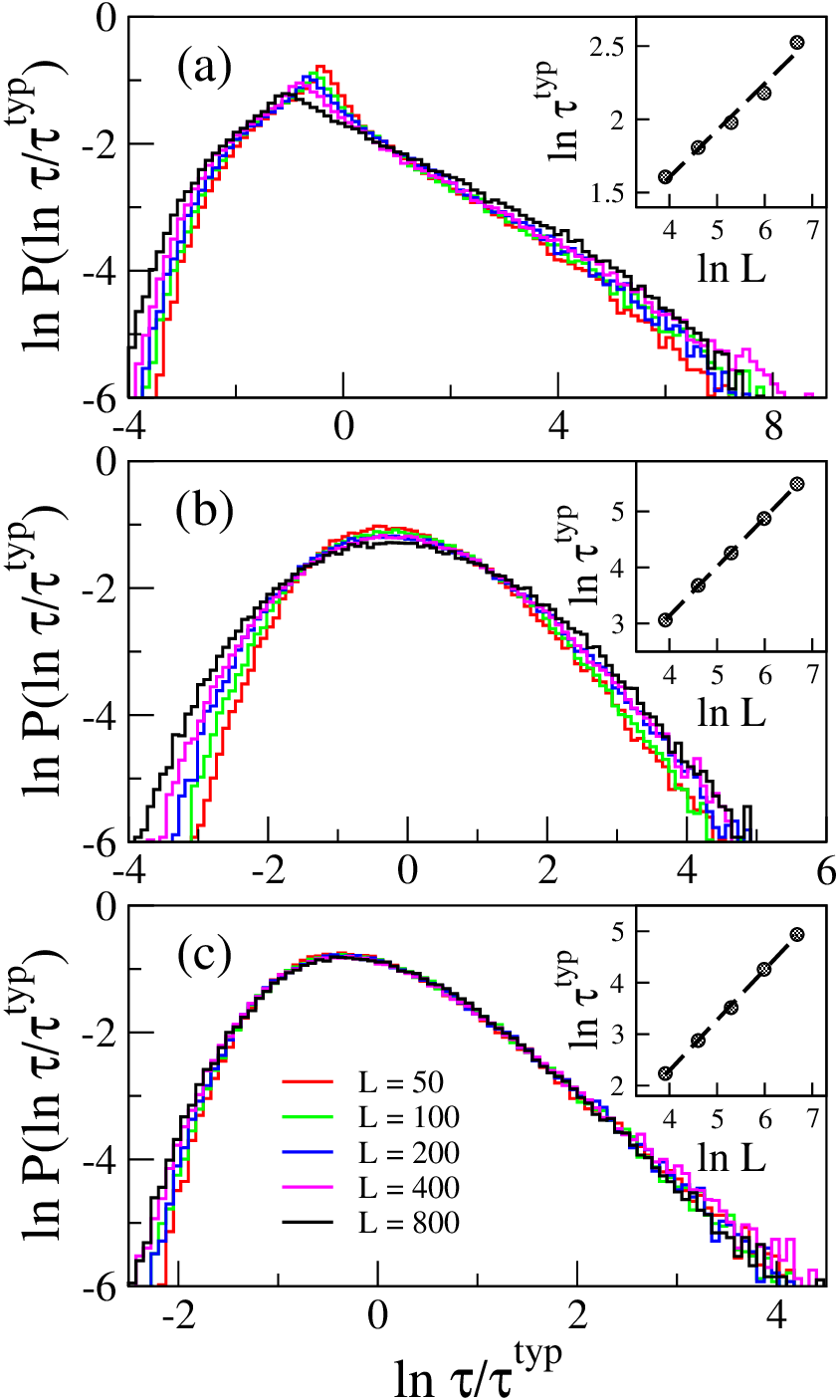}
\caption{(color online) $\ln {\cal P}(\ln \tau/\tau^{\mbox{\tiny typ}})$
as a function of $\ln \tau/\tau^{\mbox{\tiny typ}}$ for (a) $b=0.1$, (b)
$b=1$, and (c) $b=10$. In the insets the scaling
$\tau^{\mbox{\tiny typ}} \propto L^{-\mu}$ is shown together with a linear
fitting. $\tau^{\mbox{\tiny typ}} = \exp \bra \ln \tau \ket$.
We found (a) $\mu = -0.318 \pm 0.027$, (b) $\mu = -0.874\pm 0.006$, and
(c) $\mu = -0.997\pm 0.016$.}
\label{fig:fig2}
\end{center}
\end{figure}
\begin{figure}
\begin{center}
    \epsfxsize=8.4cm
    \leavevmode
    \epsffile{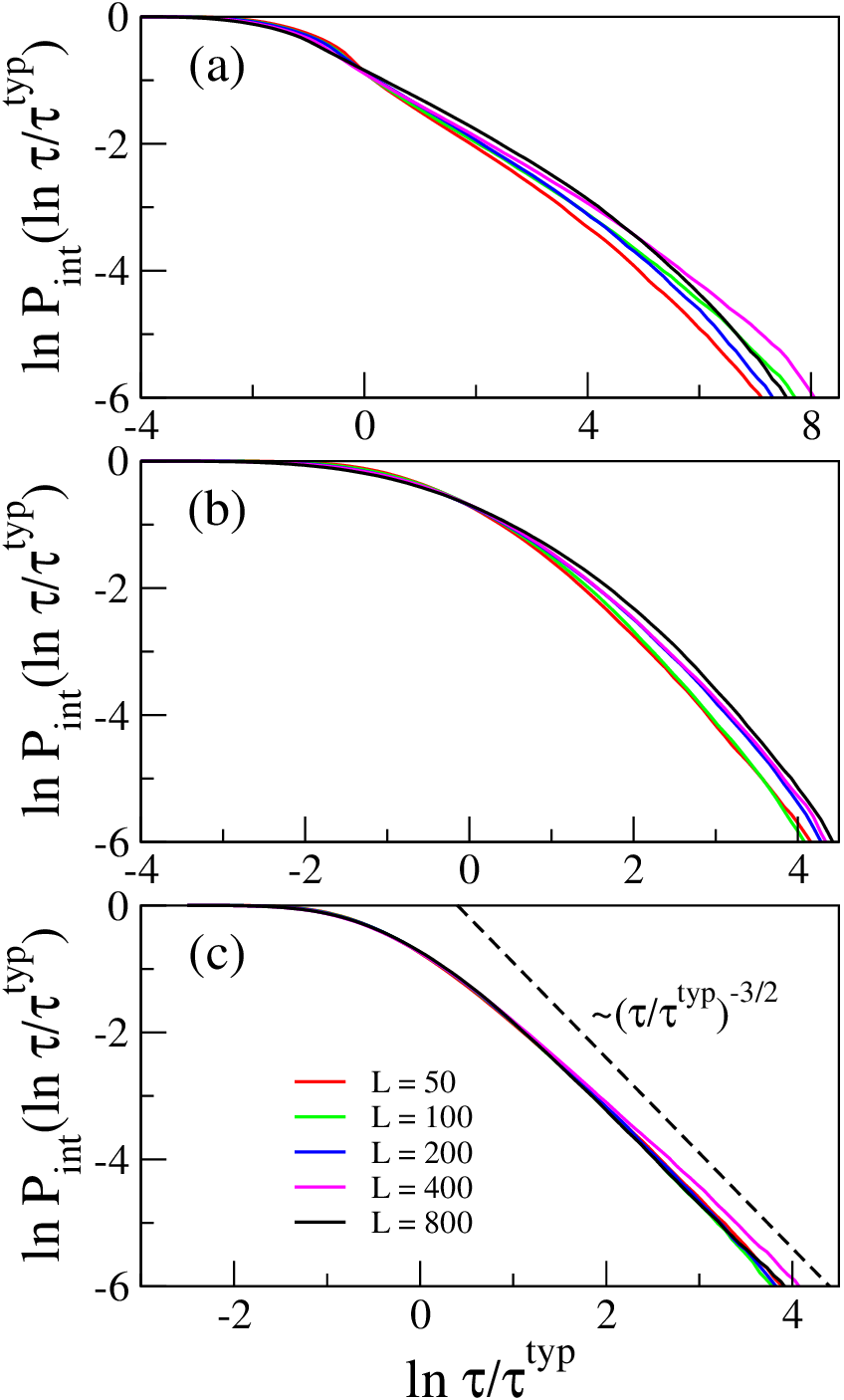}
\caption{(color online)
$\ln {\cal P}_{\mbox{\tiny int}}(\ln \tau/\tau^{\mbox{\tiny typ}})$
as a function of $\ln \tau/\tau^{\mbox{\tiny typ}}$ for (a) $b=0.1$, (b)
$b=1$, and (c) $b=10$. The thin dashed line in (c) with slope -3/2 is
plotted to guide the eye.}
\label{fig:fig3}
\end{center}
\end{figure}
\begin{figure}
\begin{center}
    \epsfxsize=8.4cm
    \leavevmode
    \epsffile{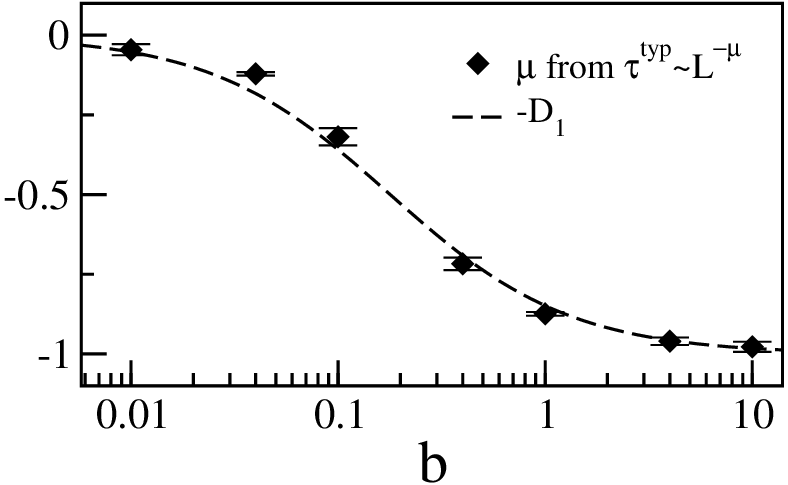}
\caption{$\mu$ as a function of $b$. $\mu$ is extracted from the
scaling $\tau^{\mbox{\tiny typ}}\propto L^{-\mu}$, see Fig.~\ref{fig:fig2}.
The dashed line is $-D_1$, where $D_1$ was obtained from Eq.~(\ref{d2}) with
$\sigma\approx 5.6$.}
\label{fig:fig6a}
\end{center}
\end{figure}

\begin{figure}
\begin{center}
    \epsfxsize=8.4cm
    \leavevmode
    \epsffile{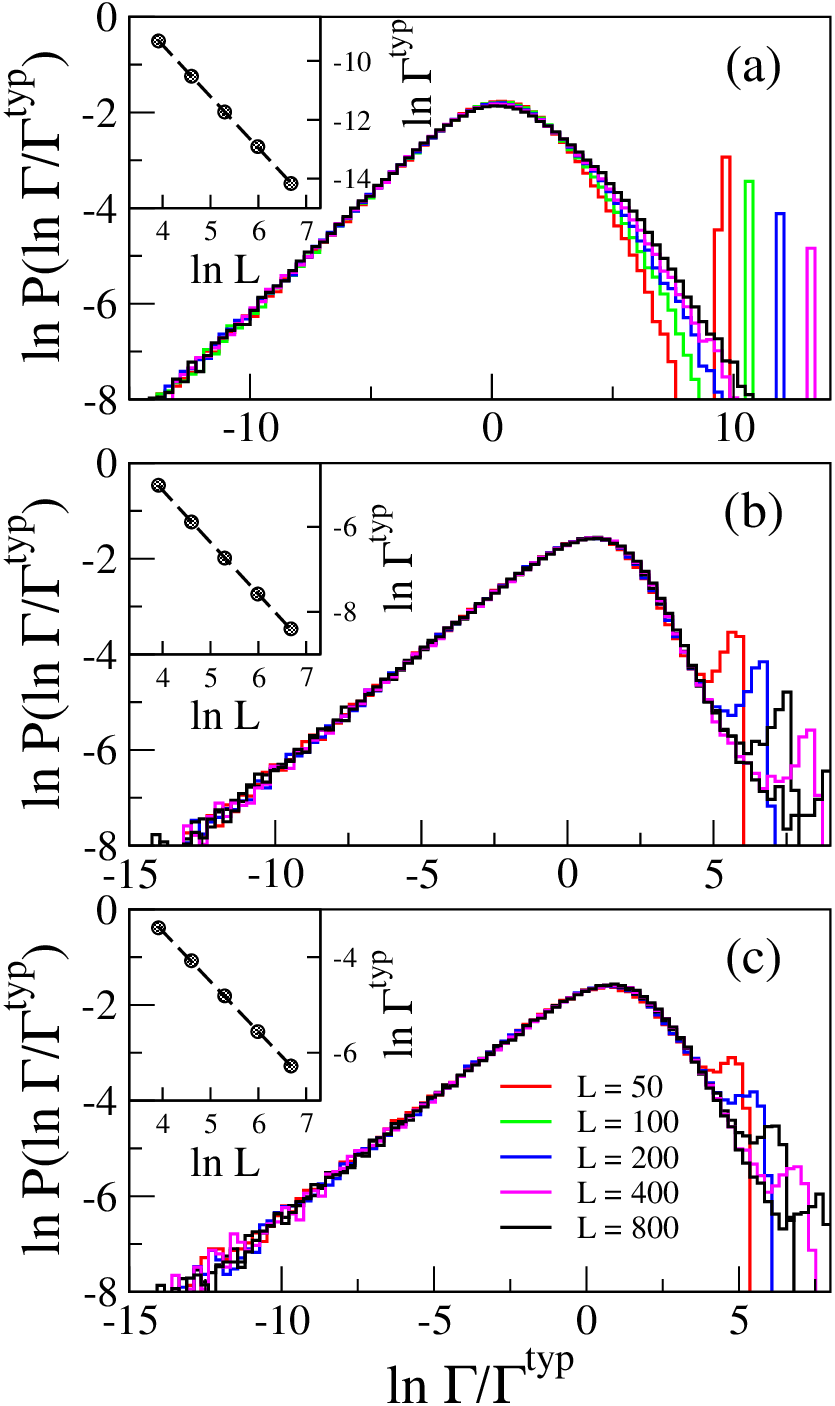}
\caption{(color online) $\ln {\cal P}(\ln \Gamma/\Gamma^{\mbox{\tiny typ}})$
as a function of $\ln \Gamma/\Gamma^{\mbox{\tiny typ}}$ for (a) $b=0.1$,
(b) $b=1$, and (c) $b=10$. In the insets the scaling
$\Gamma^{\mbox{\tiny typ}} \propto L^{-\nu}$ is shown together with a
linear fitting. $\Gamma^{\mbox{\tiny typ}} = \exp \bra \ln \Gamma \ket$.
We found (a) $\nu = 1.728 \pm 0.005$, (b) $\nu = 1.219\pm 0.007$, and
(c) $\nu = 1.048\pm 0.008$.}
\label{fig:fig4}
\end{center}
\end{figure}
\begin{figure}
\begin{center}
    \epsfxsize=8.4cm
    \leavevmode
    \epsffile{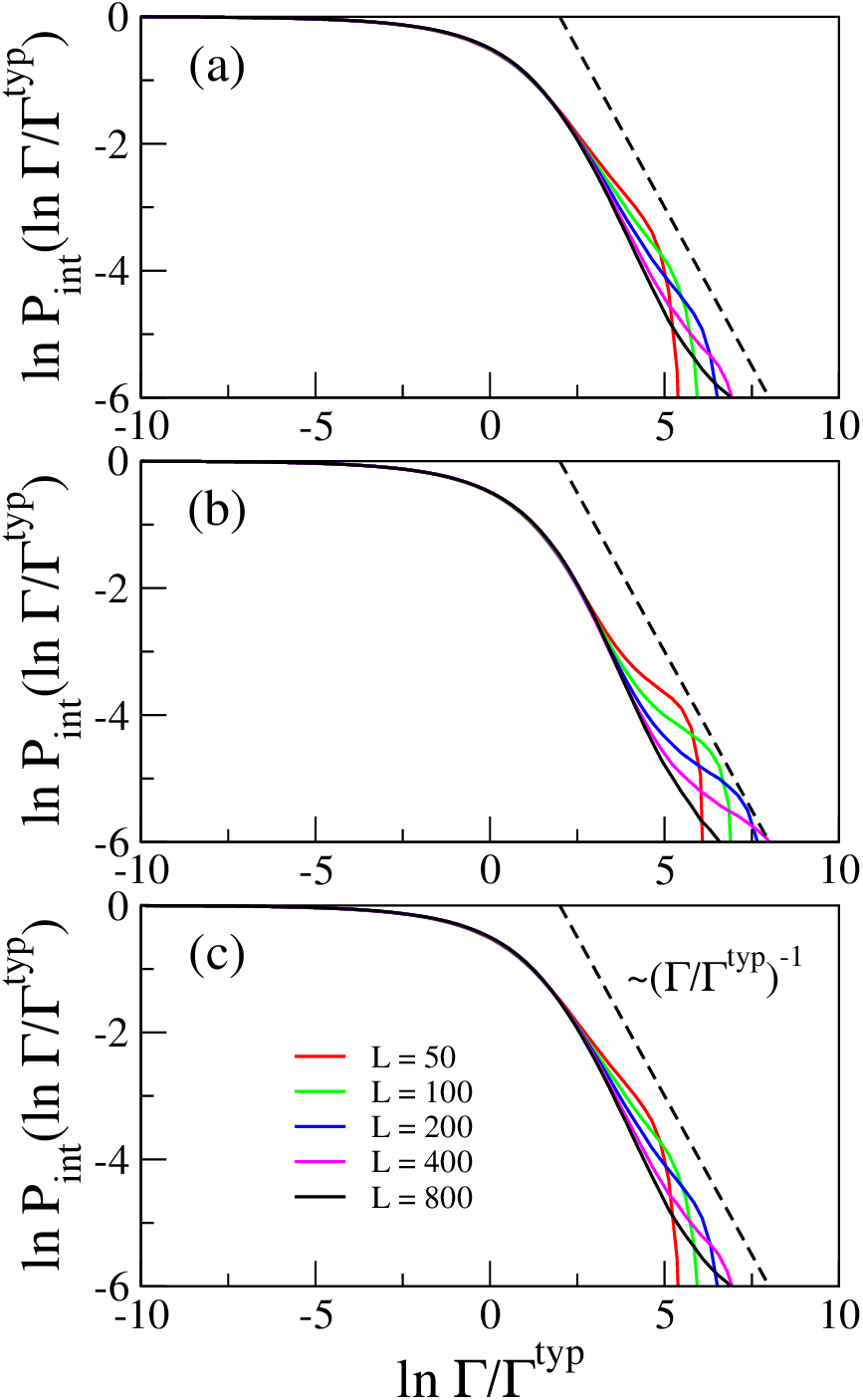}
\caption{(color online)
$\ln {\cal P}_{\mbox{\tiny int}}(\ln \Gamma/\Gamma^{\mbox{\tiny typ}})$
as a function of $\ln \Gamma/\Gamma^{\mbox{\tiny typ}}$ for (a) $b=0.1$,
(b) $b=1$, and (c) $b=10$. The thin dashed lines with slope -1 are plotted
to guide the eye.}
\label{fig:fig5}
\end{center}
\end{figure}

\section{Distribution of delay times and resonance widths}

We first have to mention that moments of the the inverse Wigner delay
time have been (numerically and analytically) shown to be related to
the multifractal dimensions of eigenfunctions as~\cite{MK05,OF05,MFME06}
\begin{equation}
\label{scaletw}
\bra\tau^{-q}\ket=L^{-qD_{q+1}},
\end{equation}
although the validity of this relation depends strongly whether the
lead is attached to a site inside or at the border of the sample.
However, in the case of the periodic PBRM model, studied here, all
sites are {\it generic} and Eq.~(\ref{scaletw}) holds well (not
shown here).

Now, let us turn to the main results of our present investigation. We
have looked at the distribution of Wigner delay times as well as that
of the resonance widths a little bit differently, compared to previous
works. First we tried to obtain a universal (i.e., $L$ independent) form
of the distribution function ${\cal P}(\ln\tau\Delta)$, where $\Delta
\propto 1/\bra \tau \ket \sim 1/L$. However, for small $b$ the relation
$\langle \tau\rangle \sim L$ does not hold and we have found a
breakdown of the scaling of ${\cal P}(\ln \tau \Delta)$ with $L$, in
contrast to the expectation according to Ref.~\onlinecite{KW02a}.

Instead of $\bra \tau \ket$ we decided to look at the typical value
of $\tau$: $\tau^{\mbox{\tiny typ}} = \exp \bra \ln \tau \ket$.
We found that ${\cal P}(\ln \tau/\tau^{\mbox{\tiny typ}})$
shows a good scaling with $L$ and a tendency to a universal distribution
(for each $b$) as $L\rightarrow \infty$, see Fig.~\ref{fig:fig2}.
In Fig.~\ref{fig:fig3} we also plot
${\cal P}_{\mbox{\tiny int}}(\ln \tau/\tau^{\mbox{\tiny typ}})$
to see the behavior of the tails of
${\cal P}(\ln \tau/\tau^{\mbox{\tiny typ}})$, where
\begin{equation}
{\cal P}_{\mbox{\tiny int}}(x) = \int_x^\infty {\cal P}(x') dx' \, .
\end{equation}
Here, only for $b=10$ we see a clear decay of the form
${\cal P}_{\mbox{\tiny int}}(\tau/\tau^{\mbox{\tiny typ}})
\sim (\tau/\tau^{\mbox{\tiny typ}})^{-3/2}$. Notice that the value of
the exponent is the same as the one found in Ref.~\onlinecite{KW02a}
for ${\cal P}_{\mbox{\tiny int}}(\tau/\bra \tau \ket)
\sim (\tau/\bra \tau \ket)^{-3/2}$ in the case of the 3D Anderson model.

In view of Eq.~(\ref{scaletw}) we may estimate the scaling behavior of
$\tau^{\mbox{\tiny typ}}$. Using a simple relation~\cite{referee} 
$\langle\ln\tau\rangle =\lim_{q\to 0}(\langle\tau^q\rangle -1)/q$, 
we obtain
\begin{equation}
\label{ttypscal}
\tau^{\mbox{\tiny typ}}\propto L^{D_1}.
\end{equation}
This is contrasted with the numerical results presented in
Fig.~\ref{fig:fig6a} where we plot $\mu$ as a function
of $b$. Here, $\mu$ is extracted from the scaling
$\tau^{\mbox{\tiny typ}}\propto L^{-\mu}$ (see the insets in
Fig.~\ref{fig:fig2}).
The data agree well with Eq.~(\ref{ttypscal}).

\begin{figure}
\begin{center}
    \epsfxsize=8.4cm
    \leavevmode
    \epsffile{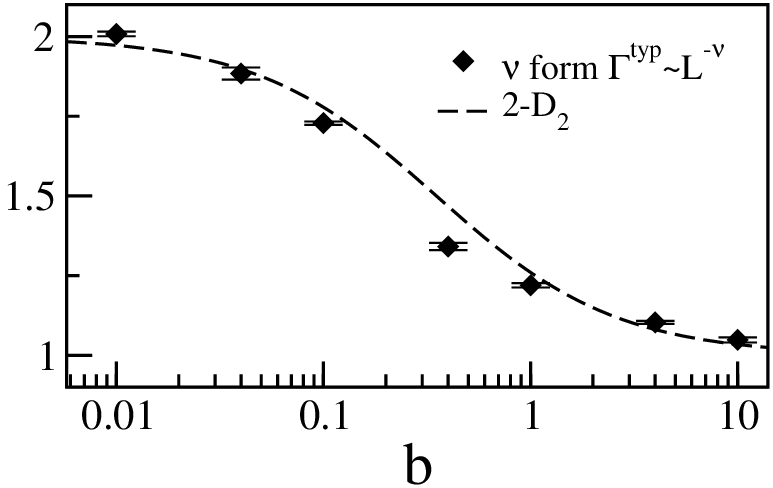}
\caption{$\nu$ as a function of $b$. $\nu$ is extracted from the scaling
$\Gamma^{\mbox{\tiny typ}} \propto L^{-\nu}$, see Fig.~\ref{fig:fig4}.
The dashed line is $2-D_2$, where $D_2$ was obtained from Eq.~(\ref{d2})
with $\sigma\approx 2.85$.}
\label{fig:fig6b}
\end{center}
\end{figure}

In the case of resonance widths, we observe an excellent scaling of
${\cal P}(\ln \Gamma/\Gamma^{\mbox{\tiny typ}})$ for all values of $b$
and the approach to a universal distribution (for each $b$) as
$L\rightarrow \infty$, see Fig.~\ref{fig:fig4}. Here,
$\Gamma^{\mbox{\tiny typ}} = \exp \bra \ln \Gamma \ket$. Also,
from Fig.~\ref{fig:fig5} we observe that as $L\rightarrow \infty$,
the integrated distribution function follows a power law decay
${\cal P}_{\mbox{\tiny int}}(\Gamma/\Gamma^{\mbox{\tiny typ}})
\sim (\Gamma/\Gamma^{\mbox{\tiny typ}})^{-1}$ for all values of $b$
in complete agreement with Refs.~\onlinecite{KW02a} and \onlinecite{WMK06}.

Finally, in Fig.~\ref{fig:fig6b} we plot $\nu$ as a function
of $b$ where $\nu$ is extracted from the scaling
$\Gamma^{\mbox{\tiny typ}} \propto L^{-\nu}$ (see the insets in
Fig.~\ref{fig:fig4}). We found that $\nu$ is well approximated by
$2-D_2$.

Notice that in the limit $b\rightarrow \infty$, where
$\nu\rightarrow1$, we recover the full random matrix limit in which
case we also have $\tau^{\mbox{\tiny typ}}\propto L\propto 1/\Delta$
and $\Gamma^{\mbox{\tiny typ}} \propto L^{-1} \propto \Delta$.
This means, for instance, that as $b\rightarrow \infty$,
${\cal P}(\Gamma/\Gamma^{\mbox{\tiny typ}}) \rightarrow
{\cal P}(\Gamma \Delta)$, where the later scaling was found in
Ref.~\onlinecite{KW02a} for the 3D Anderson model.

In Figs.~\ref{fig:fig6a} and \ref{fig:fig6b} we see that the two quantities,
$\tau^{\mbox{\tiny typ}}$ and $\Gamma^{\mbox{\tiny typ}}$, behave very
similarly and in the limit of weak multifractality, $b\gg 1$, we nicely
recover the features of the 3D Anderson transition.~\cite{KW02a}
Moreover, this similarity calls for a simple scaling law
\begin{equation}
\label{newscale}
\tau^{\mbox{\tiny typ}}\propto L^{D_1}
\qquad\mbox{and}\qquad
\Gamma^{\mbox{\tiny typ}} \propto L^{-(2-D_2)} \ .
\end{equation}

\section{Conclusions}

We have presented an analysis of the scattering properties of the
PBRM model with one lead attached to it. Especially, the system
with large conductance (weak multifractality) shows remarkable
correspondence with the properties of the standard 3D Anderson model
at criticality. This is in accordance with earlier findings.
We have also found a novel relation between the scaling of the
Shannon entropy, described by $D_1$, and the typical values of the Wigner
delay times; as well as a relation between the scaling exponent of the
participation number ${\cal N}_2$, described by $D_2$, and the
resonance widths. On the other hand the decay of
their integrated distribution functions behave just as they do for
the Anderson transition in three dimensions.

\begin{acknowledgments}
One of the authors (I.V.) acknowledges enlightening discussions with
D. V. Savin and Y. V. Fyodorov. This work was supported by the Alexander von
Humboldt Foundation, the Hungarian Research Fund (OTKA) under T42981 and
T46303, and the German-Israeli Foundation for Scientific Research and
Development.
\end{acknowledgments}

\end{document}